\documentclass{elsart}
\usepackage[english]{babel} 
\usepackage{times}
\usepackage{enumerate}
\usepackage{graphicx}
\usepackage{cite} 
\begin{document} 
\begin{frontmatter}
\title{Invasion Percolation with a Hardening Interface under Gravity} 

\author[lucas]{C.L.K. Oliveira\corauthref{cor}},
\corauth[cor]{Corresponding author.}
 \ead{zlucas@fisica.ufc.br}
\author[2]{F.K. Wittel},
\author[lucas]{J.S. Andrade, Jr.} and
\author[2]{H.J. Herrmann}

\address[lucas]{Departamento de Fisica, Universidade Federal do Ceara, Caixa Postal 6030, Campus do Pici, 60455-760 Fortaleza, Ceara, Brazil}
\address[2]{Computational Physics for Engineering Materials, IfB, ETH Zurich, Stefano-Franscini-Platz 3; CH-8093 Zurich, Switzerland} 

 \begin{abstract}
We propose a modified Invasion Percolation (IP) model to simulate the infiltration of glue into a porous medium under gravity in 2D. Initially, the medium is saturated with air and then invaded by a 
fluid that has a hardening effect taking place from the interface towards the interior by contact with the air. To take into account that interfacial hardening, we use an IP model where capillary pressures of the growth sites are increased with time. In our model, if a site stays for a certain time at interface, it becomes a hard site and cannot be invaded anymore. That represents the glue interface becoming hard due to exposition with the air. Buoyancy forces are included in this system through the Bond number which represents the competition between the hydrostatic and capillary
forces. We then compare our results with results from literature of non-hardening fluids in each regime of Bond number. We see that the invasion patterns change strongly with hardening while the non-hardening behavior remains basically not affected.
\end{abstract}
\end{frontmatter}
\newpage

\section{Introduction}

The infiltration of hardening fluids into porous media is a very interesting scientific problem with vast technological applications in many areas of knowledge. For example, in petroleum reservoir perforation, the walls of a well are cemented to insulate them from soil. In dentistry one uses adhesive bio-materials to bond resin blocks on the dentin [1,2]. Other applications are found involving wood products [3]. In all these applications, a strong and durable adhesion is required. For that, the morphological characterization of interface penetration is crucial. No matter whether a porous medium is impregnated like an antique statue or connected by adhesives, the morphology of the invaded zones determines the contact strength. On one hand, a certain invasion depth is required, for example to assure sealing, while on the other hand, the front should be rough to avoid strong gradients in mechanical properties between invaded and bulk zones.

A viscosity change of the invading fluids can be due to a chemical reaction that increases polymer chain lengths. In two-components systems, the viscosity would change globally in time and the invasion front would just \"freeze\". However in many systems the contact to air, in particular air moisture, is required for the polymer network formation. Alternatively in solvent-based systems evaporation and diffusion of solvent into the pore space and structure leads to increased viscosity. Therefore, apart from reactive resins, the contact of the invading fluid with the displaced air leads to locally increased viscosity up to hardening.

Our purpose in this work is to simulate the glue infiltration into a disordered porous medium using a 2D Invasion Percolation (IP) algorithm without trapping on a tilted square lattice under gravity. In order to take the time-dependent glue viscosity into account we propose a model that represents interfacial hardening. Hardening enters as a new parameter in our modified IP model where an interface site which stays for a long time uninvaded, hardens and cannot be invaded anymore. In the next section, we describe the hardening interface model. In Sec. 3, we show how the gravity forces can be added to an IP model. In Sec. 4, we present the results of our simulations and complete the paper with conclusions.
%%%%%%%%%%%
\section{Invasion Percolation with a Hardening Interface}\label{sec2}
The IP model was first introduced by Wilkinson and Willemsen [4] to study the immiscible two-phase displacement in porous media without gravity and chemical reactions, for the case of very slow displacements, where the capillary forces completely dominate the viscous forces. Since then, the IP paradigm has been extensively used to describe different physical systems [5-22]. In the present work, we modified the IP model to take into account the interfacial hardening of fluids.

Before describing our model, we recall the standard IP. Consider a square lattice where to each site one assigns a random number chosen between 0 and 1. In this model, each site represents a capillary pore of a disordered porous medium and each random number represents the normalized threshold capillary pressure, $p_c$, which are kept constant in time. Initially, all sites in the lattice are saturated by a defending fluid. Now another fluid invades the medium through some sites that are called growth sites or interface. If the invading fluid is injected at the left boundary, all growth sites are located there. A rough interface forms and advances into the porous medium according to the following process: at every time step, the growth site with the lowest capillary pressure is occupied by the invading fluid and removed from the list of growth sites. Due to the occupation, new sites become part of the interface, increasing the number of growth sites. An avalanche (also called burst in the literature [23-25]) can occur when the new invaded site has a pressure greater than all pressures of the invaded cluster. The occupation process is continued until the invading fluid spans the entire system, which means in our case by reaching the right boundary.

To consider the hardening of fluids in our model, the capillary pressures are not constant anymore. They increase after every avalanche by a value $\Delta$,
\begin{equation} \label{eq1}
p_c^{i;new} = p_c^{i;old} + \Delta,
\end{equation} 
where $i$ runs over all growth sites. The parameter $\Delta$ is taken to be proportional to the difference between the pressure sites, $\Delta =\alpha(p_n-p_m)$, where $p_n$ is the pressure of the new invaded site and $p_m$ is the highest pressure value in the invaded cluster. We considered different constants $\alpha$. If $p^i_c$ becomes equal or greater than unity site, $i$ hardens and cannot be invaded anymore. In this work, $p_c$ is randomly chosen from a uniform distribution. Additionally periodic boundary conditions are applied between the top and the bottom of the lattice.

In Fig. 1, we show a basic sketch of the modified IP model with fluid being injected at the left boundary (gray sites in the Figure). In this example, two avalanches occur: one, at $t = 3$ and the second at $t = 4$. Prior to the first avalanche, the IP algorithm is identical to the standard one. After the first avalanche, the new invaded site has pressure $p_n = 0.6$ while the site with the greatest pressure in the cluster has $p_m = 0.5$. In the next step the pressures of the growth sites are increased by $\alpha(0.6-0-5)$. After the second avalanche, where $p_n = 0.7$ and $p_m = 0.6$, the growth sites (sites with underlined numbers) are increased by (0:7 0:6).
%%%%%%%%%%%%%%%%%%%%%%%
\begin{figure}[htp] \centering{ \includegraphics[scale=0.65]{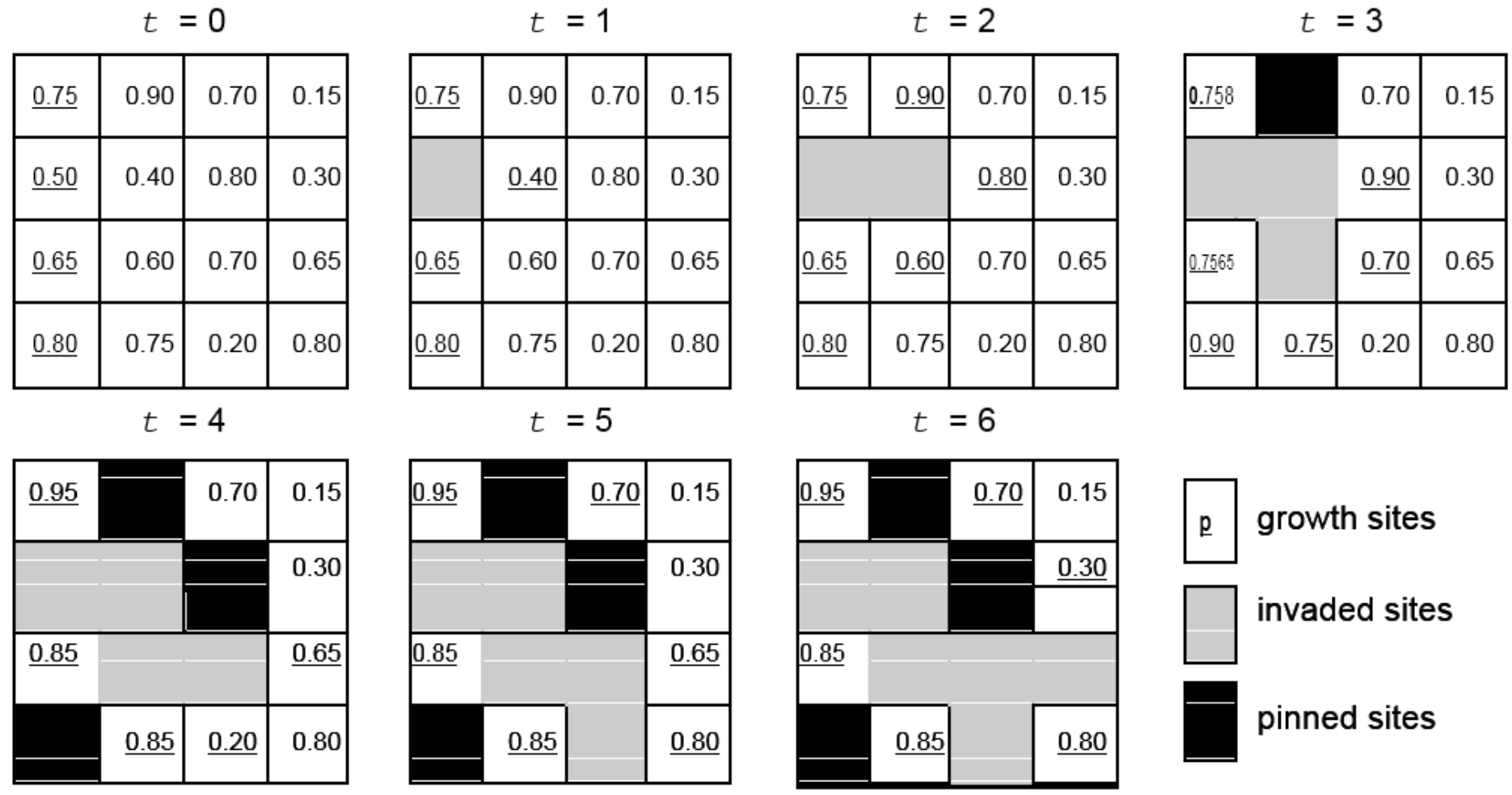} }
 \caption{\label{fig1} Scheme of IP model with hardening interface on a square lattice of $L = 4$ with infiltration starting from the left boundary for seven different iteration steps. Initially ($t = 0$), each site has a capillary pressure randomly chosen between 0 and 1. At $t = 3$ an avalanche occurs because the new invaded site has pressure ($p_n = 0.6$) greater than the greatest pressure of the cluster ($p_m = 0.5$). At this time, all sites at the interface have their pressure increased by $\alpha(0.6-0.5)$. At $t = 4$, another avalanche occurs, and now the pressures are increased again by $alpha(0.7-0.6)$. If the value of capillary pressure of a site becomes $\geq 1$, this site cannot be invaded anymore. For simplicity, $\alpha=1$ in this example. Note that only next nearest neighbors can be invaded.} 
\end{figure}

The effect of the hardening parameter $\alpha$ on the morphology is shown in Fig. 2 for the case without buoyancy forces. The picture on the left shows the pattern obtained by standard IP ($\alpha= 0$) while in the middle, the pattern for $\alpha = 0.5$ is shown, followed by the one for $\alpha= 5.0$. We observe a reduction of the invaded region because some interface sites hardened, decreasing the number of possible paths for the invasion.
\begin{figure}[htp] \centering{ \includegraphics[scale=0.65]{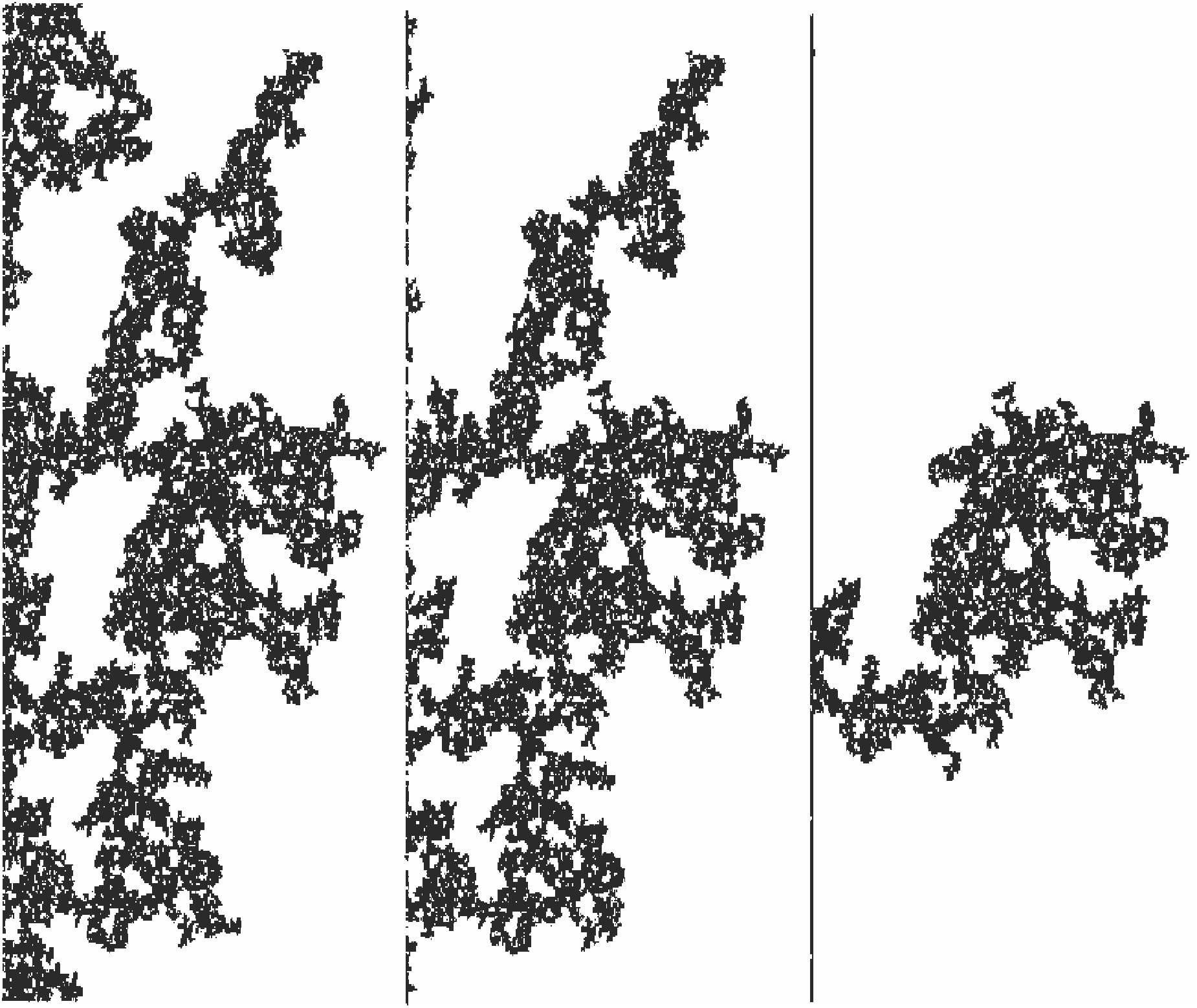} }
 \caption{\label{fig2}Three different patterns of the invading fluid at the breakthrough time for $\mathbf{Bo} = 0$ and $L = 512$. From left to right: $\alpha$= 0, 0.5, and 5.0.} 
\end{figure}
%%%%%%%%%%%%%%%%
\section{Invasion Percolation under Gravity}\label{sec3}
In the standard IP model, external forces like gravity are neglected. However, in many practical applications, for example, in a vertical displacement with fluids of different densities, it is important to consider the presence of buoyancy forces. For this purpose, Wilkinson [26] proposed a modified IP model to describe imbibition in the presence of gravity. Since then, the modified IP model has been studied extensively by other authors [27-32]. In this model, gravity causes hydrostatic pressure gradients that compete with capillary forces. That competition is expressed by the dimensionless Bond number $\mathbf{Bo}=ga^2\Delta\rho/\gamma$, where $g$ stands for the acceleration of gravity in the flow direction (here from the left to the right), $a$ denotes a typical pore size, $\Delta\rho$ describes the fluid density difference, and $\gamma$ the fluid interface tension.

Gravity is implemented in IP algorithms by adding the hydrostatic pressure gradient to the capillary pressure, $p_c$. Therefore we assign to each site $i$, the total pressure
\begin{equation} \label{eq2}
p_t^{i} = p_c^{i} + \frac{x^i\mathbf{Bo}}{2},
\end{equation} 
where $x^i$ denotes the distance from the left boundary of the lattice. Note that in our model, only the capillary pressure $p_c$ changes with the hardening effect. Therefore the hydrostatic pressure remains unaffected.

Since $\Delta\rho$ denotes the difference between the density of fluids, we can distinguish three classes of Bond numbers showing different morphologies of the front. The first one is the case without buoyancy forces ($\mathbf{Bo} = 0$) where either the system experiences no gravity or both fluids have the same density. This case is described by the standard IP algorithm. In the second case, the invaded fluid is less dense than the defending one with invading fluid being injected in the same direction as gravity. Now $\mathbf{Bo} > 0$ and gravity limits the fractal regime to a finite region, so that the front becomes self-a ne, as reported previously [32]. The last case, where a heavier fluid is injected, $\mathbf{Bo} < 0$, leads to an unstable front, also called gravity fingering [31].

The influence of the buoyancy forces in the infiltration patterns is presented in Fig. 3. The pictures on the left and in the middle, show the invasion pattern obtained for positive Bond numbers $\mathbf{Bo} = 10^-3$ and $10^-4$, respectively. In that regime, increasing $\mathbf{Bo}$ causes a more compact perimeter. In the right picture, we see the infiltration pattern for a negative Bond number ($\mathbf{Bo} = -10^-3$). In this case, there are less possible paths for the flow and the interface exhibits a fingering instability. With increasing gravity the finger straightens.

\begin{figure}[htp] \centering{ \includegraphics[scale=0.65]{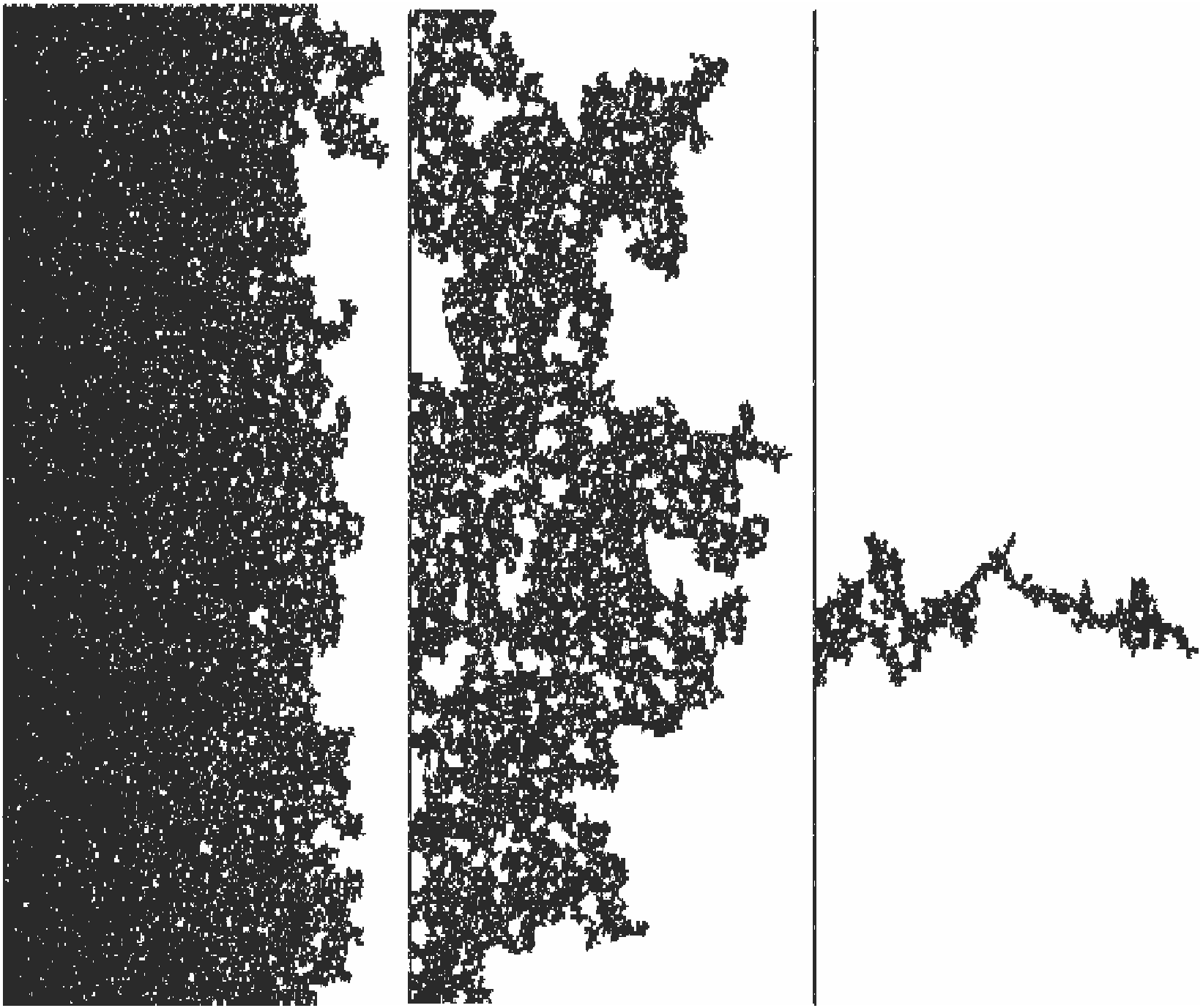} }
 \caption{\label{fig3}Three different patterns of the invading fluid at the breakthrough time for $\alpha=5.0$ and $L = 512$. From left to right: $\mathbf{Bo} = 10^-3,10^-4,$ and $-10^-3$.} 
\end{figure}
%%%%%%%%%%%%%%%%%%
\section{Results}\label{sec4}
To study the influence of hardening, we separately address the three regimes of Bond number ($\mathbf{Bo} = 0,~\mathbf{Bo} > 0,~\mathbf{Bo} < 0$). Since the invading structures are different for each regime, the analysis will be different. For $\mathbf{Bo} = 0$, we study the fractal dimension of the invaded cluster. For positive Bond numbers, we analyze the front roughness and the fractal dimension of the external perimeter and for negative Bond numbers, we explore how the percolating finger develops. All results in this work are averaged over $10^3$ realizations of lattices with size $L = 2048$.
%%%
\subsection{$\mathbf{Bo} = 0$}\label{sec41}
Zero buoyancy forces are obtained when either both fluids have the same density or the displacement takes place along a horizontal plane (neglecting the gravity). In this case, we observe for the cluster size, $M$ ,
\begin{equation} \label{eq3}
M \propto L^{d_f}
\end{equation} 
which means that the cluster is a fractal with dimension $d_f$ . We compute the fractal dimension of the cluster varying the parameter $\alpha$ of our hardening model. We see that although the invaded volume decreases with $\alpha$ (Fig. 2), the hardening effect does not affect the fractal behavior keeping the classical value of the fractal dimension $d_f = 1.89$ [33].
\begin{figure}[htp] \centering{ \includegraphics[scale=0.65]{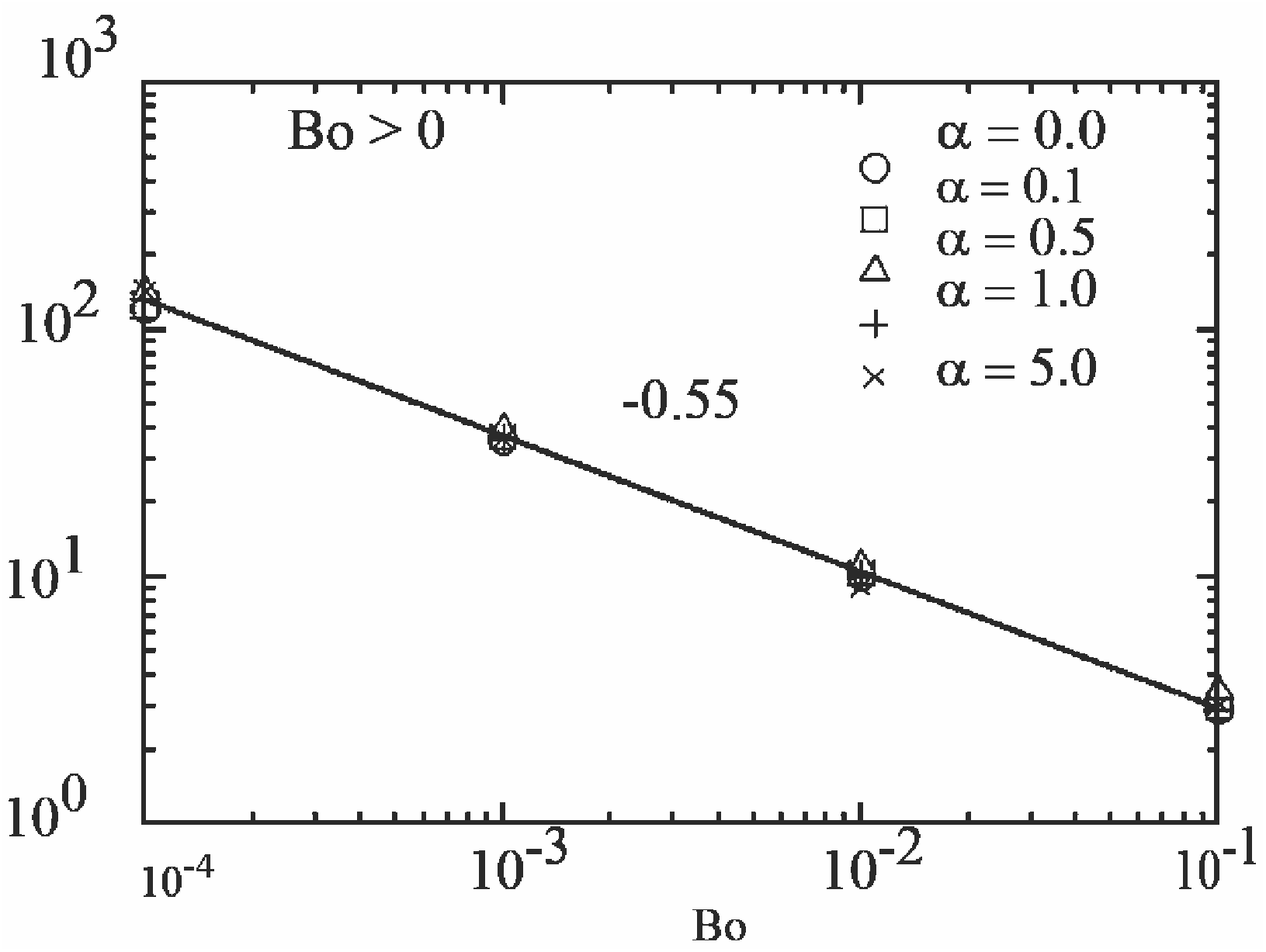} }
 \caption{\label{fig4}Front roughness at the breakthrough time vs $\mathbf{Bo}$ for different values of $\alpha$ showing that $\sigma \propto \mathbf{Bo}^\beta$, with $\beta = -0.55\pm 0.01$ for all values of $\alpha$.} 
\end{figure}
%%%%%%%
\subsection{$\mathbf{Bo} > 0$}\label{sec42}
The second case we consider is for positive Bond numbers, which represent a less-dense fluid penetrating down into a porous medium. Here gravity acts to limit the fractal regime to a finite extent. We compute the averaged front position and its front roughness defined by the standard deviation of the interface from the averaged position.

In Fig. 4, we plot the front roughness, $\sigma$, at the breakthrough time vs $\mathbf{Bo}$ in double logarithmic scale. We find that $\sigma$ follows a power-law with Bo, namely, $\sigma \propto \mathbf{Bo}^\beta$ with $\beta=-0.55\pm 0.01$. This exponent is consistent with -0.57, found by Birovljev [30]. We also compute the fractal dimension of the external perimeter via the box-counting method as being 1.39 (Fig. 5), which is the value found also in Ref. [30] through simulations. The experimentally determined value was 1.34 [30]. It is expected that the exponent approaches unity for larger values of $\mathbf{Bo}$ since the front becomes more stable.
\begin{figure}[htp] \centering{ \includegraphics[scale=0.65]{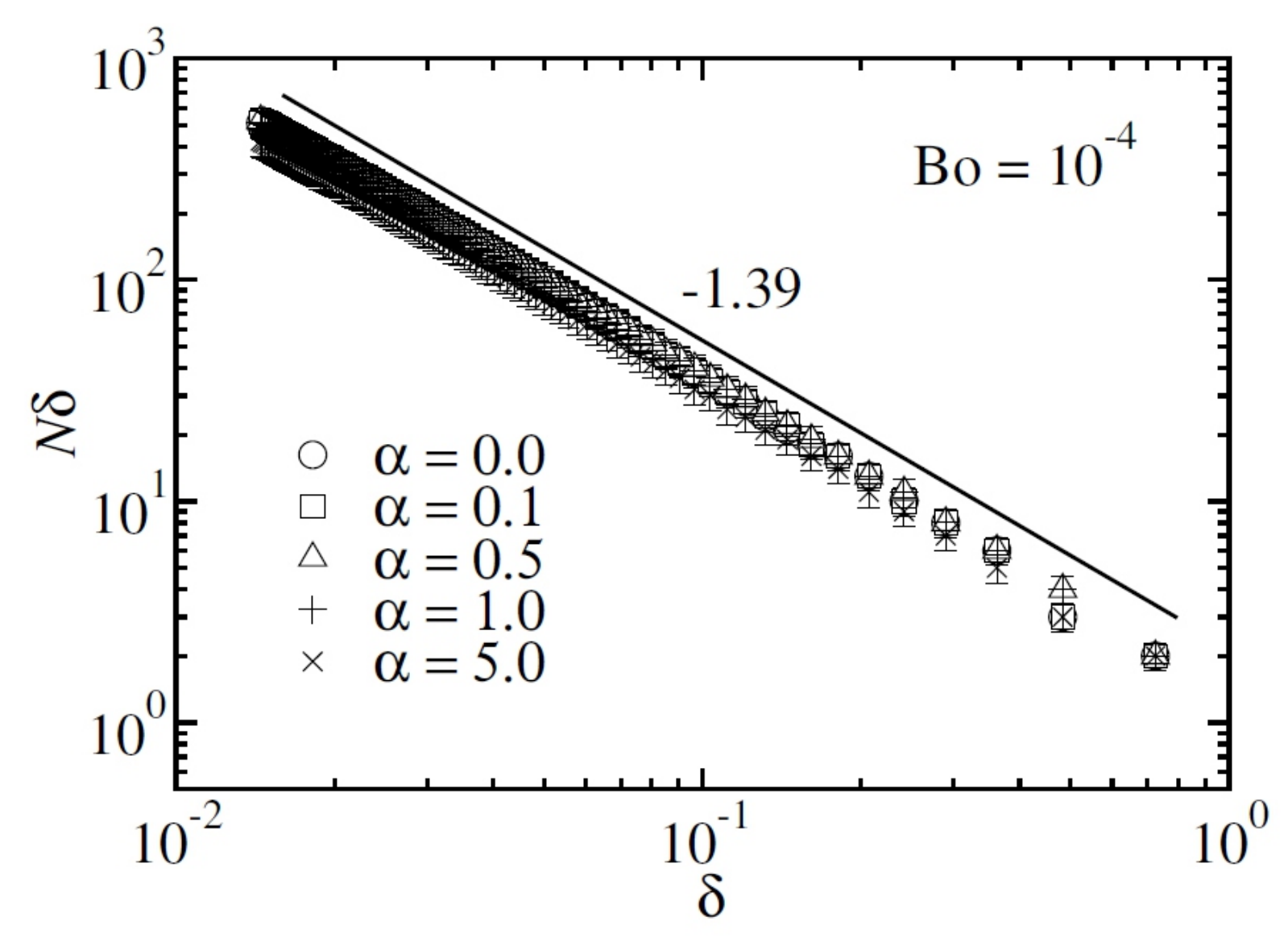} }
 \caption{\label{fig5}. Box-counting method applied to the front with different values of showing that the fractal dimension of the external perimeter for $\mathbf{Bo}=10^-4$ is 1.39.} 
\end{figure}
%%%%%%%%%%%
\subsection{$\mathbf{Bo} < 0$}\label{sec43}
For negative Bond numbers, the invading fluid is heavier than the defending one, and the front is unstable showing finger patterns. This resembles the case when glue is applied on a granular material and penetrates it by the action of gravity and capillary forces. For this case, we have studied the behavior of the finger which crosses the medium, i.e. the percolating finger. We consider only those sites of the finger that have a distance larger than $0.1L$ from the left system boundary. We compute the linear density, $S_f$ , as the number of sites of the percolating finger divided by its horizontal length. We plot $S_f$ against $\left|\mathbf{Bo}\right|$ in Fig. 6 for three different values of $\alpha$ and find that $S_f \propto \left|\mathbf{Bo}\right|^\mu$, with $\mu=-0.47\pm 0.01$. This exponent was theoretically predicted by Wilkinson, [26,32] for $alpha= 0$.

We also study how the percolating nger length and width develop as a function of the finger mass $M_f$ (Figs. 7 and 8). The percolating finger length, $h_f$, is the horizontal distance of the farthest site, belonging to the percolating finger, from the left boundary. In Fig. 7, we show the log-log of $h_f$ vs $M_f$ and observe two power-law regions. The first power-law is for small values of $\left|\mathbf{Bo}\right|$ ($\mathbf{Bo} = -10^-4$), which presents an exponent around 0.6. For intermediate values of $\left|\mathbf{Bo}\right|^\mu$, we can observe during the development of the finger a crossover from the first power-law to another one with exponent close to unity. The second power-law region increases with $\left|\mathbf{Bo}\right|$ and dominates the finger behavior for $\mathbf{Bo} = -10^-1$. The proportionality is expected because for $\left|\mathbf{Bo}\right|$ large enough, the finger grows only in the gravitational direction.
\begin{figure}[htp] \centering{ \includegraphics[scale=0.65]{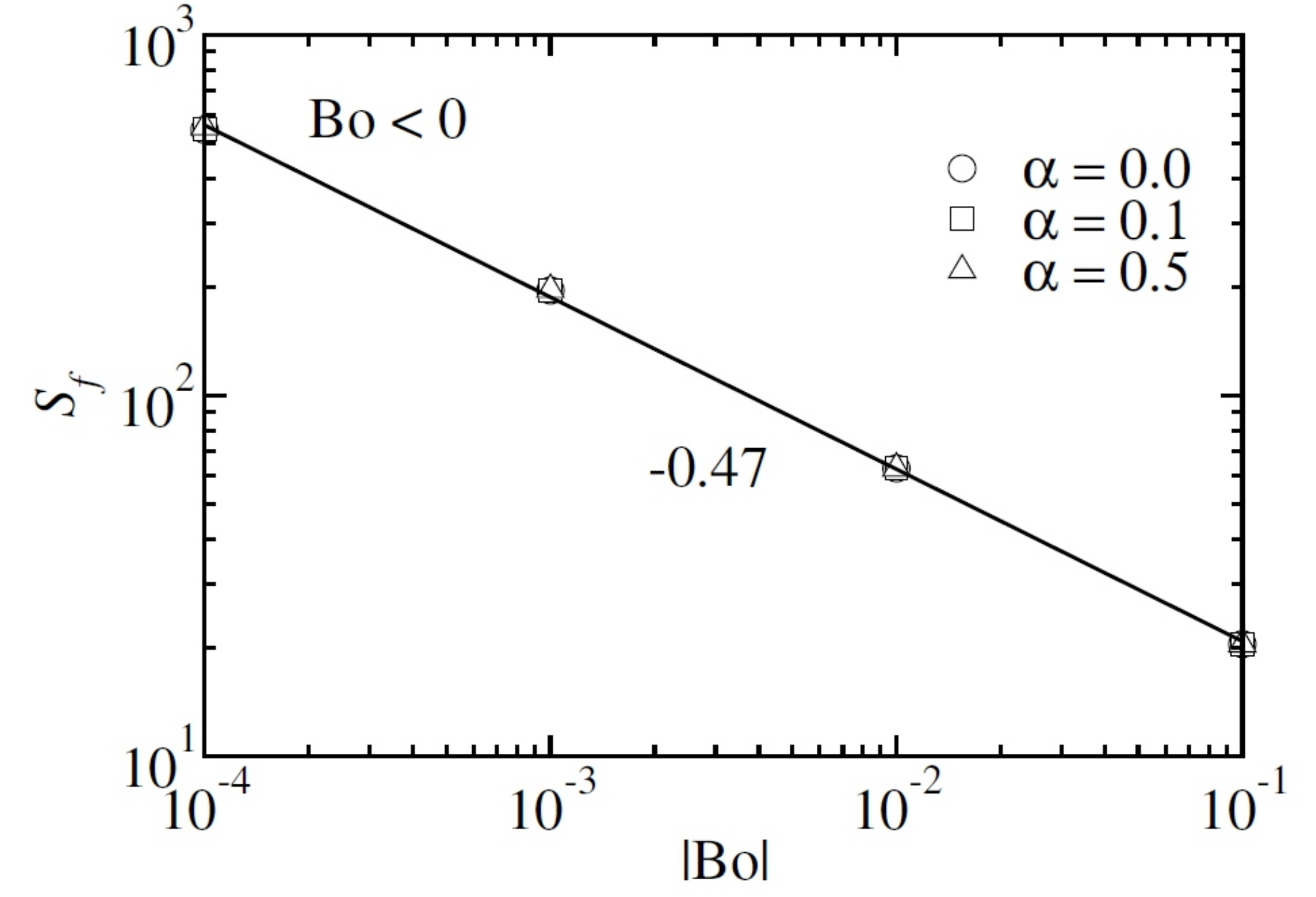} }
 \caption{\label{fig6}.	Linear density of the percolating   finger $S_f$  vs $\left|\mathbf{Bo}\right|$ for different values of showing that $S_f \propto \left|\mathbf{Bo}\right|$ , with $\mu=-0.47\pm 0.01$.} 
\end{figure}

We also study the development of the percolating finger width. We apply a linear regression to the distance of the sites which belong to the percolating finger. The finger width, $w_f$, is then defined as the standard deviation of these distances to the regression. The invasion dynamics is captured by repeating this procedure for every increment since the regression line changes as the finger grows. The finger width grows according to a power-law with exponent $\lambda_w = 0.52\pm 0.05$ for $\mathbf{Bo} = -10^-4$. When the finger grows enough, the width seems to saturate to a constant value, as we can see in Fig. 8.
\begin{figure}[htp] \centering{ \includegraphics[scale=0.65]{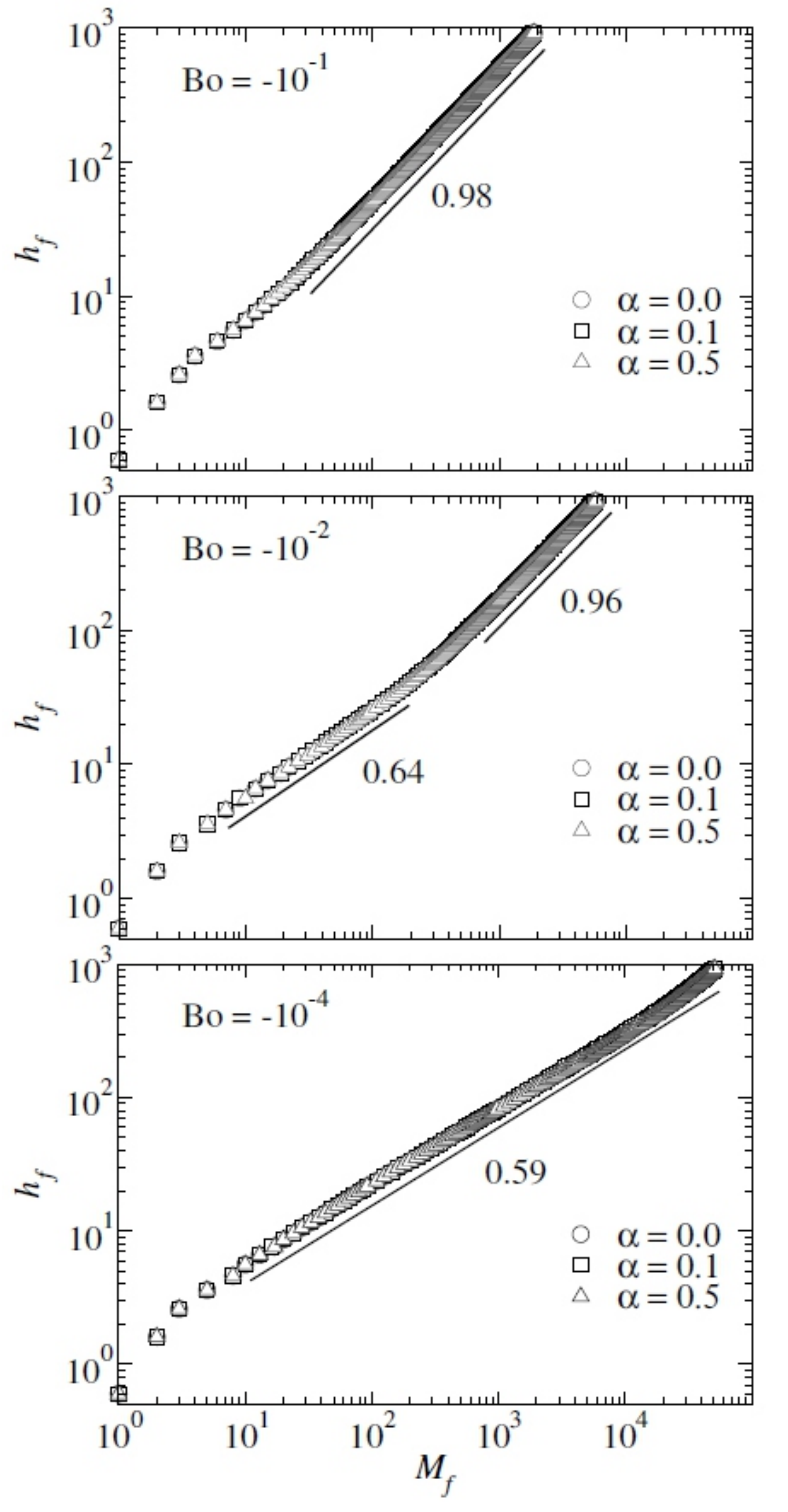} }
 \caption{\label{fig7}.	Percolating finger length vs finger mass for different values of and $\mathbf{Bo}$. Two power-law regimes $h_f \propto M_f^{\lambda_h}$ are observed: where $\lambda_h = 0.98\pm 0.02$ for high values of $\left|\mathbf{Bo}\right|$ and $\lambda_h = 0.59\pm 0.03$ for low values of $\left|\mathbf{Bo}\right|$. For intermediate values of $\left|\mathbf{Bo}\right|$ a crossover separates these two regimes.	} 
\end{figure}

We have also considered a different model to represent the hardening. Instead of increasing the capillary pressure by a parameter $\Delta$ which depends on the site pressures at every avalanche time, we incrementally increase it at every time iteration and consider the parameter $\Delta$ as a constant during the simulation using various values of $\Delta$ (=$10^-4,~10^-3,$ and $10^-2$). Although the hardening effect is stronger in this way, our results do not change in any regime of Bond number.
\begin{figure}[htp] \centering{ \includegraphics[scale=0.65]{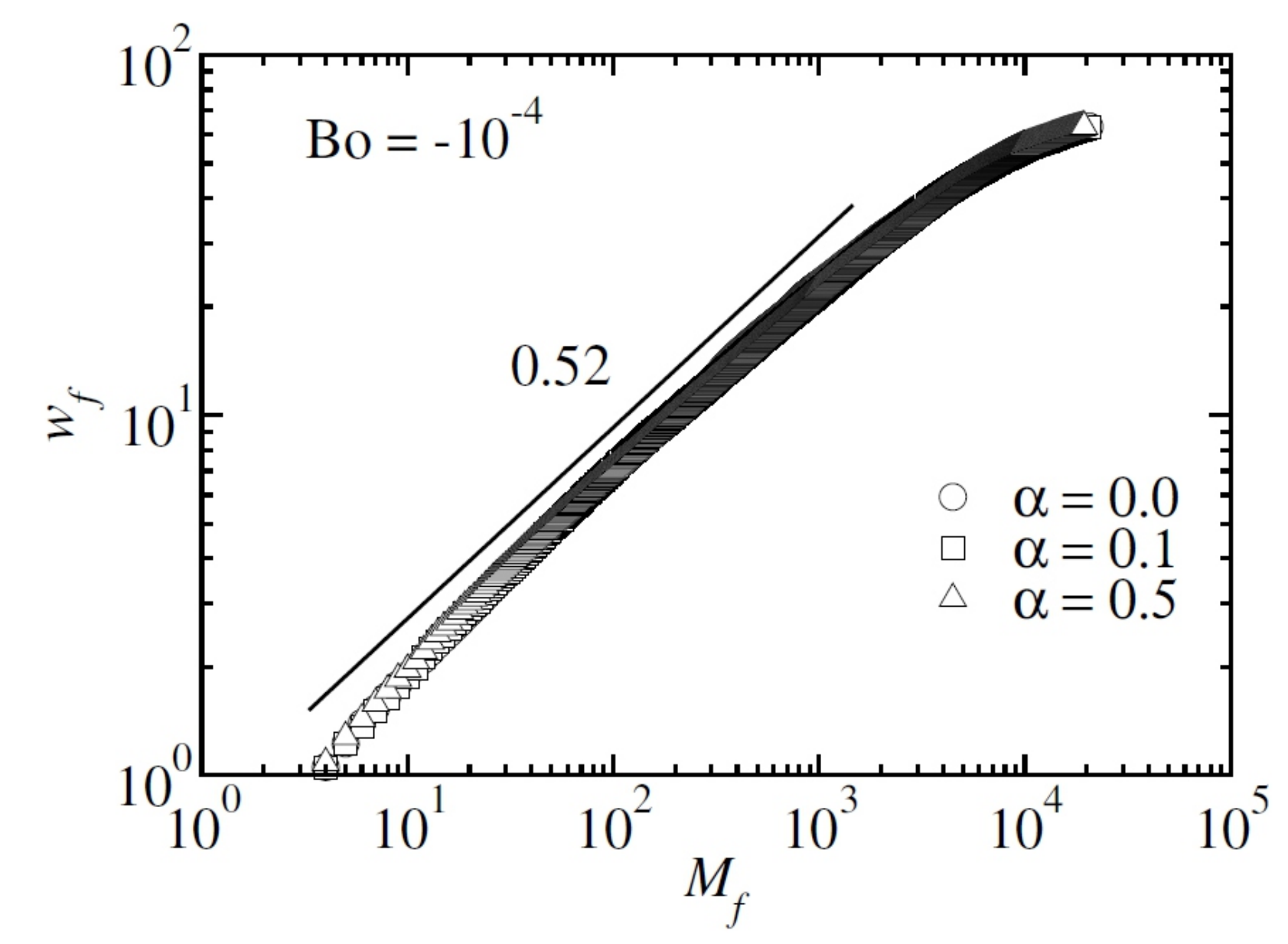} }
 \caption{\label{fig8}.	Percolating  finger width vs finger mass for different values of $\alpha$ showing that $w_f \propto M_f^{\lambda_w}$, with $\lambda_w = 0.52\pm 0.05$ for $\mathbf{Bo} = -10^-4$.} 
\end{figure}
%%%%%%%%%%%%%%%%%%
\section{Conclusions}\label{sec5}
In this paper, we proposed a modified IP model to simulate the flow of a hardening fluid into porous media under gravity. As an IP model, each site in the system has a capillary pressure randomly chosen between 0 and 1. To describe hardening we increase these pressures by a parameter $\Delta$ after every avalanche. Only the pressure of the interface sites is increased by that parameter. If a site reaches a capillary pressure equal or greater than unity, it hardens and cannot be invaded anymore. The results were averaged over $10^3$ realizations for a tilted square lattice with size $L = 2048$.

We considered different regimes of Bond numbers ($\mathbf{Bo} = 0, ~\mathbf{Bo} > 0$ and $\mathbf{Bo} < 0$) and studied the front patterns for each regime differently. For $\mathbf{Bo} = 0$, we computed the fractal dimension of the invaded cluster. For positive Bond numbers, we analyzed the front roughness and the fractal dimension of the external perimeter and for negative Bond numbers, we looked at the growth of the linear density of the percolating finger with $\left|\mathbf{Bo}\right|$. We then compared our results with known results from literature of non-hardening fluids in each regime of Bond number. We see that the invasion patterns change strongly with hardening.

We calculated for the first time the percolating finger length and width for negative values of Bond number. According to our results, the percolating finger length grows with its mass obeying two power-law regions. The first power-law is for small values of $\left|\mathbf{Bo}\right|$ ($\mathbf{Bo} = -10^-4$), which presents an exponent around 0.6. For intermediate values of $\left|\mathbf{Bo}\right|$ we can observe during the development of the finger a crossover from the first power-law to another one with exponent close to unity. This second power-law increases with $\left|\mathbf{Bo}\right|$ and dominates the finger behavior for $\mathbf{Bo} = -10^-1$. The finger width is then computed as the standard deviation of these sites. The finger width grows according to a power-law with exponent $\lambda_w =0.52\pm 0.05$ for $\mathbf{Bo} = -10^-4$. When the finger grows sufficiently the width seems to saturate to a constant value.
%%%%%%%%%%%%%%%%
\section{Acknowledgment} 
We thank S. Roux for enlightening discussions. This work was supported by CNPq, CAPES, FINEP, FUNCAP, Petrobras, FUNCAP/CNPq (Pronex), and the National Institute of Science and Technology for Complex Systems. Also the support of the Swiss National Science Foundation (SNF) under Grant No. 116052 is acknowledged.
%%%%%%%%%%%%%%%%
\section*{References}
\begin{enumerate}[ {[}1{]} ]
\item	S. K. Moura, J. F. F. Santos and R. Y. Ballester, Braz. Dent. J. \textbf{17}, 179 (2006). 
\item	V. de P. A. Saboia, S. K. Saito and L. A. F. Pimenta, Pesqui. Odontol. Bras. \textbf{14}, 340 (2000). 
\item	C. R. Frihart, J. ASTM Internacional \textbf{2}, JAI12952 (2005). 
\item	D. Wilkinson and J. F. Willemsen, J. Phys. A \textbf{16}, 3365 (1983). 
\item	R. Lenormand and C. Zarcone, Phys. Rev. Lett. \textbf{54}, 2226 (1985). 
\item	M. Cieplak, A. Maritan and J. R. Banavar, Phys. Rev. Lett. \textbf{76}, 3754 (1996). 
\item	M. Porto, S. Havlin, S. Schwarzer and A. Bunde, Phys. Rev. Lett. \textbf{79}, 4060 (1997). 
\item	M. Hashemi, M. Sahimi and B. Dabir, Phys. Rev. Lett. \textbf{80}, 3248 (1998). 
\item	M. Sahimi, M. Hashemi and J. Ghassemzadeh, Physica A \textbf{260}, 231 (1998). 
\item	A. P. Sheppard, M. A. Knackstedt, W. V. Pinczewski and M. Sahimi, J. Phys. A \textbf{32}, L521 (1999). 
\item	M. A. Knackstedt, M. Sahimi and A. P. Sheppard, Phys. Rev. E \textbf{61}, 4920 (2000). 
\item	S. Zapperi, A. A. Moreira and J. S. Andrade, Jr., Phys. Rev. Lett. \textbf{86}, 3622 (2001). 
\item	R. Dobrin and P. M. Duxbury, Phys. Rev. Lett. \textbf{86}, 5076 (2001). 
\item	M. A. Knackstedt, S. J. Marrink, A. P. Sheppard, W. V. Pinczewski and M. Sahimi, Transp. Porous Media \textbf{44}, 465 (2001). 
\item		M. A. Knackstedt, M. Sahimi and A. P. Sheppard, Phys. Rev. E \textbf{65}, 035101 (2002). 
\item	A. D. Araujo, J. S. Andrade, Jr. and H. J. Herrmann, Phys. Rev. E \textbf{70}, 066150 (2004). 
\item	A. D. Araujo, T. F. Vasconcelos, A. A. Moreira, L. S. Lucena and J. S. Andrade, Jr., Phys. Rev. E \textbf{72}, 041404 (2005). 
\item	A. Gabrielli and G. Caldarelli, Phys. Rev. Lett. \textbf{98}, 208701 (2007). 
\item	A. D. Araujo, M. C. Romeu, A. A. Moreira, R. F. S. Andrade and J. S. Andrade, Jr., Phys. Rev. E \textbf{77}, 041410 (2008). 
\item	J. Shao, S. Havlin and H. E. Stanley, Phys. Rev. Lett. \textbf{103}, 018701 (2009). 
\item	E. Fehr, J. S. Andrade, Jr., S. D. da Cunha, L. R. da Silva, H. J. Herrmann, D. Kadau, C. F. Moukarzel and E. A. Oliveira, J. Stat. Mech. P09007 (2009). 
\item	J. S. Andrade, Jr., E. A. Oliveira, A. A. Moreira and H. J. Herrmann, Phys. Rev. Lett. \textbf{103}, 225503 (2009). 
\item	S. Roux and E. Guyon, J. Phys. A \textbf{22}, 3693 (1989). 
\item	L. Furuberg, J. Feder, A. Aharony and T. J  ssang, Phys. Rev. Lett. \textbf{61}, 2117 (1988). 
\item	A. Dougherty and N. Carle, Phys. Rev. E \textbf{58}, 2889 (1998). 
\item	D. Wilkinson, Phys. Rev. A \textbf{30}, 520 (1984). 
\item	J. F. Gouyet, M. Rosso and B. Sapoval, Phys. Rev. B \textbf{37}, 1832 (1988). 
\item	D. Or, Adv. Water Res. \textbf{31}, 1129 (2008). 
\item	G. L voll, Y. Meheust, K. J. Mal y, E. Aker and J. Schmittbuhl, Energy \textbf{30}, 861 (2005). 
\item	A. Birovljev, L. Furuberg, J. Feder, T. J ssang, K. J. Mal y and A. Aharony, Phys. Rev. Lett. \textbf{67}, 584 (1991). 
\item	V. Frette, J. Feder, T. J  ssang and P. Meakin, Phys. Rev. Lett. \textbf{68}, 3164 (1992). 
\item	M. Chaouche, N. Rakotomalala, D. Salin, B. Xu and Y. C. Yortsos, Phys. Rev. E \textbf{49}, 4133 (1994). 
\item	J. Feder, Fractals (Plenum, New York, 1988), p. 135. 
\end{enumerate}

\end{document}